\documentclass[12pt,journal,letterpaper,onecolumn]{IEEEtran}

\usepackage{amsmath}
\usepackage{amssymb}
\usepackage{mathptmx}
\usepackage{graphicx}
\usepackage{times}
\usepackage{wrapfig}
\usepackage{algorithm}
\usepackage{algorithmic}

\newtheorem{theorem}{Theorem}[section]
\newtheorem{lemma}[theorem]{Lemma}

\newtheorem{corollary}[theorem]{Corollary}
\newtheorem{problem}[theorem]{Problem}
\newtheorem{definition}{Definition}[section]

\title{Discrete Laplace-Beltrami Operator Determines Discrete Riemannian Metric}
\author{Xianfeng David Gu$^\dag$,\thanks{$^\dag$Xianfeng David Gu, Department of Computer Science, Stony Brook University, Stony Brook, NY 11794, gu@cs.sunysb.edu.}
\and Ren Guo$^\ddagger$,\thanks{$^\ddagger$Ren Guo, School of Mathematics, University of Minnesota, Minneapolis, MN 55455,  guoxx170@math.umn.edu.}
\and  Feng Luo$^\ast$,\thanks{$^\ast$Feng Luo, Department of Mathematics, Rutgers University, Piscataway, NJ 08854,  fluo@math.rutgers.edu.}
\and Wei Zeng$^\S$ \thanks{$^\S$Wei Zeng, Department of Computer Science, Stony Brook University, Stony Brook, NY 11794, zengwei@cs.sunysb.edu.}
}

\begin{document}
\maketitle

\begin{abstract}
The Laplace-Beltrami operator of a smooth Riemannian manifold is
determined by the Riemannian metric. Conversely, the heat kernel
constructed from its eigenvalues and eigenfunctions determines the
Riemannian metric. This work proves the analogy on Euclidean
polyhedral surfaces (triangle meshes), that the discrete
Laplace-Beltrami operator and the discrete Riemannian metric (unique
up to a scaling) are mutually determined by each other.

Given an Euclidean polyhedral surface, its Riemannian metric is
represented as edge lengths, satisfying triangle inequalities on all
faces. The Laplace-Beltrami operator is formulated using the
cotangent formula, where the edge weight is defined as the sum of
the cotangent of angles against the edge. We prove that the edge
lengths can be determined by the edge weights unique up to a scaling
using the variational approach.

First, we show that the space of all possible metrics of a
polyhedral surface is convex. Then, we construct a special energy
defined on the metric space, such that the gradient of the energy
equals to the edge weights. Third, we show the Hessian matrix of the
energy is positive definite, restricted on the tangent space of the
metric space, therefore the energy is convex. Finally, by the fact
that the parameter on a convex domain and the gradient of a convex
function defined on the domain have one-to-one correspondence, we
show the edge weights determines the polyhedral metric unique up to
a scaling.

The constructive proof leads to a computational algorithm that finds
the unique metric on a topological triangle mesh from a discrete
Laplace-Beltrami operator matrix.
\end{abstract}

\section{Introduction}

Laplace-Beltrami operator plays a fundamental role in Riemannian
geometry \cite{Rosenberg98}. Discrete Laplace-Beltrami operators on
triangulated surface meshes span the entire spectrum of geometry
processing applications, including mesh parameterization,
segmentation, reconstruction, compression, re-meshing and so on
\cite{Levy06Laplace,Sorkine06Differential,Zhang09Sprectral}.
Laplace-Beltrami operator is determined by the Riemannian metric.
The heat kernel can be constructed from the eigenvalues and
eigenfunctions of the Laplace-Beltrami operator, conversely, it
fully determines the Riemannian metric (unique
up to a scaling). In this work, we prove the
discrete analogy to this fundamental fact, that the discrete
Laplace-Beltrami operator and the discrete Riemannian metric are
mutually determined by each other.

\smallskip
{\textbf{Related Works }}
In real applications, a smooth metric surface is usually represented
as a triangulated mesh. The manifold heat kernel is estimated from
the discrete Laplace operator. The most well-known and widely-used
discrete formulation of Laplace operator over triangulated meshes is
the so-called \emph{cotangent scheme}, which was originally
introduced in \cite{Dodziuk78Sprectral,Pinkall93MinimalSurface}. Xu
\cite{Xu04Convergence} proposed several simple discretization
schemes of Laplace operators over triangulated surfaces, and
established the theoretical analysis on convergence. Wardetzky et
al. \cite{08Nofreelunch} proved the theoretical limitation that the
discrete Laplacians cannot satisfy all natural properties, thus,
explained the diversity of existing discrete Laplace operators. A
family of operations were presented by extending more natural
properties into the existing operators. Reuter et
al. \cite{Reuter06ShapeDNA} computed a discrete Laplace operator
using the finite element method, and exploited the isometry
invariance of the Laplace operator as shape fingerprint for object
comparison. Belkin et al. \cite{Belkin08MeshLaplace} proposed the
first discrete Laplacian that pointwise converges to the true
Laplacian as the input mesh approximates a smooth manifold better.
Tamal et al. \cite{Tamal10Spectra} employed this mesh Laplacian and
provided the first convergence to relate the discrete spectrum with
the true spectrum, and studied the stability and robustness of the
discrete approximation of Laplace spectra. The eigenfunctions of
Laplace-Beltrami operator have been applied for global intrinsic
symmetry detection in \cite{OvsjanikovSG08}. Heat Kernel Signature
was proposed in \cite{SunOG09}, which is concise and characterizes the
shape up to isometry.

\smallskip
\textbf{Our Results }In this work, we prove that the discrete
Laplace-Beltrami operator based on the cotangent
scheme \cite{Dodziuk78Sprectral,Pinkall93MinimalSurface} is
determined by the discrete Riemannian metric, and also determines
the metric unique up to a scaling. The proof is using the
variational approach, which leads to a practical algorithm to
compute a Riemannian metric from a prescribed Laplace-Beltrami
operator.

\smallskip
\textbf{Paper Outline } In Section \ref{sec:overview}, we briefly overview the
fundamental theorem of smooth heat kernel and our theoretical claims
of discrete case. We clarify the simplest case, one triangle mesh
in Section \ref{sec:triangle} first; then turn to the more general
Euclidean polyhedral surfaces in Section \ref{sec:polyhedral}. Finally,
in Section \ref{sec:experiment}, we present a variational algorithm
to compute the unique Riemannian metric from from a Laplace-Beltrami
matrix. The numerical experiments on different topological triangle
meshes support the theoretic results.

\section{Preliminaries and Proof Overview}
\label{sec:overview}

\subsection{Smooth Case}
\label{sec:smooth}
Suppose $(M,\mathbf{g})$ is a complete
Riemannian manifold, $\mathbf{g}$ is the Riemannian metric. $\Delta$ is the
Laplace-Beltrami operator. The eigenvalues $\{\lambda_n\}$ and
eigenfunctions $\{\phi_n\}$ of $\Delta$ are
\[
    \Delta \phi_n = -\lambda_n \phi_n,
\]
where $\phi_n$ is normalized to be orthonormal in $L^2(M)$. The
spectrum is given by
\[
0=\lambda_0 < \lambda_1 \le \lambda_2 \le \cdots, ~~~~\lambda_n \to
\infty.
\]

Then there is a heat kernel $K(x,y,t) \in C^\infty(M\times M \times
\mathbb{R}^+)$, such that
\[
    K(x,y,t) = \sum_{n=0}^\infty e^{-\lambda_nt} \phi_n(x)\phi_n(y).
\]
Heat kernel reflects all the information of the Riemannian metric
$\mathbf{g}$. The details of the following theorem can be found in
\cite{SunOG09}.

\smallskip
\begin{theorem} Let $f: (M_1,\mathbf{g}_1)\to (M_2,\mathbf{g}_2)$ be a diffeomorphism
between two Riemannian manifolds. If $f$ is an isometry, then
\begin{equation}
    K_1(x,y,t) = K_2( f(x),f(y),t),~\forall x,y \in M,~t >0.
    \label{eqn:heat_kernel_2}
\end{equation}
Conversely, if $f$ is a surjective map, and
Eqn. (\ref{eqn:heat_kernel_2}) holds, then $f$ is an isometry.
\end{theorem}

\subsection{Discrete Case}
\label{sec:discrete}
In this work, we focus on discrete surfaces, namely polyhedral
surface. For example, a triangle mesh is piecewise linearly embedded in
$\mathbb{R}^3$.

\smallskip
\begin{definition} [{Polyhedral Surface}] An Euclidean polyhedral surface is a triple
$(S,T,\mathbf{d})$, where $S$ is a closed surface, $T$ is a triangulation of
$S$ and $\mathbf{d}$ is a metric on $S$ whose restriction to each triangle is
isometric to an Euclidean triangle.
\end{definition}

The well-known cotangent edge weight
\cite{Dodziuk78Sprectral,Pinkall93MinimalSurface} on an Euclidean
polyhedral surface is defined as follows: 

\smallskip
\begin{definition}[{Cotangent Edge Weight}]
\label{def:edge_weight} Suppose $[v_i,v_j]$ is a boundary edge of
$M$, $[v_i,v_j] \in
\partial M$, then $[v_i,v_j]$ is associated with one triangle $[v_i,v_j,v_k]$, the angle
against $[v_i,v_j]$ at the vertex $v_k$ is $\alpha$, then the weight
of $[v_i,v_j]$ is given by  $w_{ij} = \frac{1}{2}\cot \alpha$.
Otherwise, if $[v_i,v_j]$ is an interior edge, the two angles
against it are $\alpha,\beta$, then the weight is $w_{ij} =
\frac{1}{2}(\cot \alpha + \cot \beta)$.
\end{definition}

The discrete Laplace-Beltrami operator is constructed from the cotangent
edge weight.

\smallskip
\begin{definition}[{Discrete Laplace Matrix}]
\label{def:laplace} The discrete Laplace matrix $L=(L_{ij})$ for an
Euclidean polyhedral surface is given by
\[
    L_{ij} =
    \left\{
    \begin{array}{ll}
    -w_{ij}& i \neq j\\
    \sum_k w_{ik} & i = j\\
    \end{array}
    \right..
\]
\end{definition}

Because $L$ is symmetric, it can be decomposed as
\begin{equation}
    L = \Phi \Lambda \Phi^T
    \label{eqn:laplace}
\end{equation}
where $\Lambda=diag(\lambda_0,\lambda_1,\cdots,\lambda_n)$,
$0=\lambda_0 < \lambda_1 \le \lambda_2 \le \cdots \le \lambda_n$, are the
eigenvalues of $L$, and $\Phi=(\phi_0|\phi_1|\phi_2|\cdots|\phi_n)$, $L
\phi_i = \lambda_i \phi_i$, are the orthonormal eigenvectors, such that
$\phi_i^T\phi_j = \delta_{ij}$.

\smallskip
\begin{definition}[{Discrete Heat Kernel}]
 The discrete heat kernel is defined as follows:
 \begin{equation}
    K(t)=\Phi exp( -\Lambda t ) \Phi^T.
    \label{eqn:heat_kernel}
 \end{equation}
\end{definition}

The \textbf{Main Theorem}, called \emph{Global Rigidity Theorem}, in this work is as follows:

\smallskip
\begin{theorem}
\label{thm:main} Suppose two Euclidean polyhedral surfaces
$(S,T,\mathbf{d_1})$ and $(S,T,\mathbf{d_2})$ are given,
\[
    L_1=L_2,
\]
if and only if $\mathbf{d_1}$ and $\mathbf{d_2}$ differ by a
scaling.
\end{theorem}

\smallskip
\begin{corollary}
\label{cor:main} Suppose two Euclidean polyhedral surfaces
$(S,T,\mathbf{d_1})$ and $(S,T,\mathbf{d_2})$ are given,
\[
    K_1(t)=K_2(t), \forall t > 0,
\]
if and only if $\mathbf{d_1}$ and $\mathbf{d_2}$ differ by a
scaling.
\end{corollary}

\smallskip
\begin{proof}
Note that,
\[
    \frac{dK(t)}{dt}|_{t=0} = -L.
\]
Therefore, the discrete Laplace matrix and the discrete heat kernel
mutually determine each other.
\end{proof}

\subsection{Proof Overview for Main Theorem \ref{thm:main}}
The main idea for the proof is as follows. We fix the connectivity
of the polyhedral surface $(S,T)$. Suppose the edge set of $(S,T)$
is sorted as $E=\{e_1, e_2,\cdots, e_m\}$, where $m=|E|$ number of
edges, the face set is denoted as $F$. A triangle $[v_i,v_j,v_k]\in
F$ is also denoted as $\{i,j,k\}\in F$.

By definition, an Euclidean polyhedral metric on $(S,T)$ is given by
its edge length function $d:E\to \mathbb{R}^+$. We denote a metric
as $\mathbf{d}=(d_1,d_2,\cdots,d_m)$, where $d_i=d(e_i)$ is the length of edge $e_i$. Let
\[
    E_{\mathbf{d}}(2) = \{(d_1,d_2,d_3)| d_i + d_j > d_k\}
\]
be the space of all Euclidean triangles parameterized by the edge
lengths, where $\{i,j,k\}$ is a cyclic permutation of $\{1,2,3\}$.
In this work, for convenience, we use $u=(u_1,u_2,\cdots,u_m)$ to
represent the metric, where $u_k = \frac{1}{2} d_k^2$.

\smallskip
\begin{definition}[{Admissible Metric Space}]
\label{def:metric_space}Given a triangulated surface $(S,K)$, the
admissible metric space is defined as
\[
  \Omega_u = \{(u_1,u_2,u_3\cdots,u_m)| \sum_{k=1}^m u_k = m,  (\sqrt{u_i},\sqrt{u_j},\sqrt{u_k})\in E_{\mathbf{d}}(2), \forall \{i,j,k\}\in F\}.
\]
\end{definition}

We show that $\Omega_u$ is a convex domain in $\mathbb{R}^{m}$.

\smallskip
\begin{definition}[{Energy}]
\label{def:energy} An energy $E:\Omega_u \to \mathbb{R}$ is defined
as:
\begin{equation}
E(u_1,u_2 \cdots, u_m ) =
\int^{(u_1,u_2\cdots,u_m)}_{(1,1,\cdots,1)} \sum_{k=1}^m w_k(\mu)
d\mu_k, \label{eqn:energy}
\end{equation}
where $w_k(\mu)$ is the cotangent weight on the edge $e_k$ determined by
the metric $\mu$.
\end{definition}

Next we show this energy is convex in Lemma \ref{lem:convexity_energy}. According to the following
lemma, the gradient of the energy $\nabla E({\mathbf{d}}):\Omega\to
\mathbb{R}^m$
\[
    \nabla E: (u_1,u_2\cdots,u_m)\to (w_1,w_2,\cdots w_m)
\]
is an embedding. Namely the metric is determined by the edge weight
unique up to a scaling.

\smallskip
\begin{lemma}Suppose $\Omega \subset \mathbb{R}^n$ is an open convex domain in $\mathbb{R}^n$,
$E: \Omega \to \mathbb{R}$ is a strictly convex function with
positive definite Hessian matrix, then $\nabla E:\Omega \to
\mathbb{R}^n$ is a smooth embedding. \label{lem:embedding}
\end{lemma}

\smallskip
\begin{proof}
If $\mathbf{p}\neq \mathbf{q}$ in $\Omega$, let $\gamma(t) =
(1-t)\mathbf{p} + t\mathbf{q} \in \Omega$ for all $t\in [0,1]$. Then
$f(t)=E(\gamma(t)):[0,1]\to \mathbb{R}$ is a strictly convex
function, so that
\[
    \frac{d f(t)}{dt} = \nabla E|_{\gamma(t)} \cdot (\mathbf{q}-\mathbf{p}).
\]
Because
\[
    \frac{d^2 f(t)}{dt^2} = (\mathbf{q}-\mathbf{p})^T
    H|_{\gamma(t)}(\mathbf{q}-\mathbf{p}) > 0,
\]
$\frac{d f(0)}{dt} \neq \frac{d f(1)}{dt}$, therefore
\[
    \nabla E(\mathbf{p}) \cdot (\mathbf{q}-\mathbf{p}) \neq
    \nabla E(\mathbf{q}) \cdot (\mathbf{q}-\mathbf{p}).
\]
This means $\nabla E(\mathbf{p}) \neq \nabla E(\mathbf{q})$,
therefore $\nabla E$ is injective.

On the other hand, the Jacobi matrix of $\nabla E$ is the Hessian
matrix of $E$, which is positive definite. It follows that $\nabla
E:\Omega \to \mathbb{R}^n$ is a smooth embedding.
\end{proof}

From the discrete Laplace-Beltrami operator (Eqn. (\ref{eqn:laplace}))
or the heat kernel (Eqn. (\ref{eqn:heat_kernel})), we can compute all
the cotangent edge weights, then because the edge weight determines
the metric, we attain the Main Theorem \ref{thm:main}.

\section{Euclidean Triangle}
\label{sec:triangle}
In this section, we show the proof for the simplest case, a Euclidean triangle; in the next
section, we generalize the proof to all types of triangle meshes.

\begin{figure*}
\centering
\begin{tabular}{c}
\includegraphics[height=1.50in]{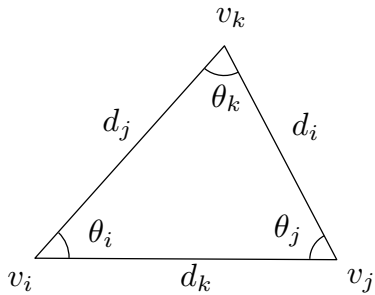} \\
\end{tabular}
\caption{An Euclidean triangle.} \label{fig:triangle}
\end{figure*}

Given a triangle $\{i,j,k\}$, three corner angles denoted by $\{\theta_i,\theta_j,\theta_k\}$,
three edge lengths denoted by $\{d_i,d_j,d_k\}$, as shown in Fig. \ref{fig:triangle}. In this case, the problem is trivial. Given
$(w_i,w_j,w_k)=(\cot\theta_i,\cot\theta_j,\cot\theta_k)$, we can
compute $(\theta_i,\theta_j,\theta_k)$ by taking the $\arctan$
function. Then the normalized edge lengths are given by
\[
    (d_i,d_j,d_k) =
    \frac{3}{\sin\theta_i+\sin\theta_j+\sin\theta_k}(\sin\theta_i,\sin\theta_j,\sin\theta_k).
\]

Although this approach is direct and simple, it can not be
generalized to more complicated polyhedral surfaces. In the
following, we use a different approach, which can be generalized to
all polyhedral surfaces.

\smallskip
\begin{lemma}Suppose an Euclidean triangle is with angles
$\{\theta_i,\theta_j,\theta_k\}$ and edge lengths $\{d_i,d_j,d_k\}$,
angles are treated as the functions of the edge lengths $\theta_i(d_i,d_j,d_k)$, then
\begin{equation}
\frac{\partial \theta_i}{\partial d_i} = \frac{d_i}{2A}
\end{equation}
and
\begin{equation}
\frac{\partial \theta_i}{\partial d_j} =
-\frac{d_i}{2A}\cos\theta_k,
\end{equation}
where $A$ is the area of the triangle.
\end{lemma}

\smallskip
\begin{proof}
According to Euclidean cosine law,
\begin{equation}
\cos \theta_i = \frac{d_j^2 + d_k^2 - d_i^2}{2d_j d_k},
\end{equation}
we take derivative on both sides with respective to $d_i$,
\[
    -\sin\theta_i \frac{\partial \theta_i}{\partial d_i} =\frac{-2d_i}{2d_jd_k}
\]
\begin{equation}
\begin{split}
\frac{\partial \theta_i}{\partial d_i} &=
\frac{d_i}{d_jd_k\sin\theta_i} = \frac{d_i}{2A}
\end{split}
\end{equation}
where $A = \frac{1}{2}d_jd_k \sin\theta_i$ is the area of the
triangle. Similarly,
\[
\frac{\partial}{\partial
d_j}(d_j^2+d_k^2-d_i^2)=\frac{\partial}{\partial d_j}
(2d_jd_k\cos\theta_i)
\]
\[
 2 d_j = 2d_k \cos\theta_i -
2d_jd_k\sin\theta_i \frac{\partial \theta_i}{\partial d_j}
\]
\[
2A  \frac{\partial \theta_i}{\partial d_j} =  d_k\cos\theta_i - d_j
= -d_i \cos\theta_k
\]
We get
\[
\frac{\partial \theta_i}{\partial d_j} = -\frac{d_i\cos\theta_k}{2A}.
\]
\end{proof}

\smallskip
\begin{lemma}
\label{lem:symmetry} In an Euclidean triangle,  let $u_i =
\frac{1}{2}d_i^2$ and $u_j=\frac{1}{2}d_j^2$ then
\begin{equation}
    \frac{\partial \cot \theta_i }{\partial u_j} =  \frac{\partial \cot \theta_j }{\partial u_i}
\end{equation}
\end{lemma}

\smallskip
\begin{proof}
\begin{equation}
\begin{split}
    \frac{\partial \cot \theta_i }{\partial u_j}
    &=\frac{1}{d_j}\frac{\partial \cot \theta_i }{\partial d_j}=-\frac{1}{d_j}\frac{1}{\sin^2\theta_i}\frac{\partial\theta_i}{\partial
d_j}
    =\frac{1}{d_j}\frac{1}{\sin^2\theta_i}\frac{d_i\cos\theta_k}{2A} =\frac{d_i^2 }{\sin^2 \theta_i} \frac{\cos\theta_k}{2A d_id_j}\\
    &=\frac{4R^2}{2A} \frac{\cos\theta_k}{d_id_j}
\end{split}
\label{eqn:symmetry}
\end{equation}
where $R$ is the radius of the circum circle of the triangle. The
righthand side of Eqn. (\ref{eqn:symmetry}) is symmetric with respect to the indices $i$ and $j$.
\end{proof}

\smallskip
\begin{corollary}
\label{cor:closed_1_form} The differential form
\begin{equation}
    \omega = \cot \theta_i du_i + \cot \theta_j du_j + \cot \theta_k
    du_k
    \label{eqn:1_form}
\end{equation}
is a closed 1-form.
\end{corollary}

\smallskip
\begin{proof}
By the above Lemma \ref{lem:symmetry} regarding symmetry,
\[
\begin{split}
d\omega &= (\frac{\partial\cot\theta_j}{\partial
u_i}-\frac{\partial\cot\theta_i}{\partial u_j}) du_i \wedge
du_j+(\frac{\partial\cot\theta_k}{\partial
u_j}-\frac{\partial\cot\theta_j}{\partial u_k}) du_j \wedge du_k
\\&+(\frac{\partial\cot\theta_i}{\partial u_k}-\frac{\partial
\cot\theta_k}{\partial u_i}) du_k
\wedge du_i\\
& = 0.
\end{split}
\]
\end{proof}

\smallskip
\begin{definition}[{Admissible Metric Space}]
Let $u_i=\frac{1}{2}d_i^2$, the admissible metric space is defined
as
\[
\Omega_u := \{(u_i,u_j,u_k)|(\sqrt{u_i},\sqrt{u_j},\sqrt{u_k})\in
E_{\mathbf{d}}(2),~u_i+u_j+u_k = 3\}
\]
\end{definition}

\smallskip
\begin{lemma} The admissible metric space $\Omega_u$ is a convex domain in $\mathbb{R}^3$.
\label{lem:convexity_metric_space}
\end{lemma}

\smallskip
\begin{proof}
Suppose $(u_i,u_j,u_k)\in \Omega_u$ and
$(\tilde{u}_i,\tilde{u}_j,\tilde{u}_k)\in \Omega_u$, then from
$\sqrt{u_i} + \sqrt{u_j} > \sqrt{u_k}$, we get $u_i + u_j +
2\sqrt{u_iu_j} > u_k$. Define $(u_i^\lambda,u_j^\lambda,u_k^\lambda)
= \lambda (u_i,u_j,u_k) + (1-\lambda)
(\tilde{u}_i,\tilde{u}_j,\tilde{u}_k)$, where $0<\lambda <1$. Then
\[
\begin{split}
u_i^\lambda u_j^\lambda&=(\lambda u_i +
(1-\lambda)\tilde{u}_i)(\lambda u_j +
(1-\lambda)\tilde{u}_j)\\
&=\lambda^2 u_i u_j + (1-\lambda)^2 \tilde{u}_i\tilde{u}_j +
\lambda(1-\lambda) (u_i\tilde{u}_j+u_j\tilde{u}_i)\\
&\ge\lambda^2 u_i u_j + (1-\lambda)^2 \tilde{u}_i\tilde{u}_j +
2\lambda(1-\lambda)\sqrt{u_iu_j\tilde{u}_i\tilde{u}_j}\\
&=(\lambda\sqrt{u_iu_j}+(1-\lambda)\sqrt{\tilde{u}_i\tilde{u}_j})^2
\end{split}
\]
It follows
\[
\begin{split}
u_i^\lambda + u_j^\lambda + 2\sqrt{u_i^\lambda u_j^\lambda}
&\ge \lambda(u_i + u_j + 2\sqrt{u_iu_j}) + (1-\lambda)(\tilde{u}_i +
\tilde{u}_j + 2\sqrt{\tilde{u}_i\tilde{u}_j})\\&>\lambda u_k + (1-\lambda) \tilde{u}_k =  u_k^\lambda
\end{split}
\]
This shows $(u_i^\lambda,u_j^\lambda,u_k^\lambda)\in \Omega_u$.
\end{proof}

Similarly, we define the edge weight space as follows.

\smallskip
\begin{definition}[{Edge Weight Space}]
The edge weights of an Euclidean triangle form the edge weight space
\[
\Omega_\theta = \{(\cot\theta_i, \cot \theta_j, \cot \theta_k)| 0 <
\theta_i,\theta_j,\theta_k < \pi, \theta_i + \theta_j + \theta_k =
\pi \}.
\]
\end{definition}

Note that,
\[
    \cot\theta_k = -\cot(\theta_i+\theta_j) = \frac{1-\cot\theta_i\cot\theta_j}{\cot\theta_i + \cot\theta_j}.
\]

\begin{figure*}
\centering
\begin{tabular}{c}
\includegraphics[height=1.50in]{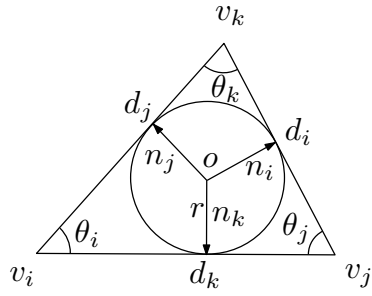} \\
\end{tabular}
\caption{The geometric interpretation of the Hessian matrix. The
in circle of the triangle is centered at $O$, with radius $r$. The perpendiculars $n_i$,
$n_j$ and $n_k$ are from the incenter of the triangle and orthogonal
to the edge $e_i$, $e_j$ and $e_k$ respectively.}
\label{fig:hessian}
\end{figure*}

\smallskip
\begin{lemma}
The energy $E: \Omega_u \to \mathbb{R}$
\begin{equation}
\label{eqn:energy} E(u_i,u_j,u_k) = \int_{(1,1,1)}^{(u_i,u_j,u_k)}
\cot\theta_i d\tau_i + \cot \theta_j d\tau_j + \cot \theta_k d\tau_k
\end{equation}
is well defined on the admissible metric space $\Omega_u$ and is
convex. \label{lem:convexity_energy}
\end{lemma}

\smallskip
\begin{proof}
According to Corollary \ref{cor:closed_1_form}, the differential form is
closed. Furthermore, the admissible metric space $\Omega_u$ is a
simply connected domain. The differential form is exact, therefore,
the integration is path independent, and the energy function is well
defined.

Then we compute the Hessian matrix of the energy,
\[
H=-\frac{2R^2}{A} \left[
\begin{array}{ccc}
\frac{1}{d_i^2}&-\frac{\cos\theta_k}{d_id_j}&-\frac{\cos\theta_j}{d_id_k}\\
-\frac{\cos\theta_k}{d_jd_i}&\frac{1}{d_j^2}&-\frac{\cos\theta_i}{d_jd_k}\\
-\frac{\cos\theta_j}{d_kd_i}&-\frac{\cos\theta_i}{d_kd_j}&\frac{1}{d_k^2}\\
\end{array}
\right] = -\frac{2R^2}{A} \left[
\begin{array}{ccc}
(\mathbf{\eta}_i,\mathbf{\eta}_i)&(\mathbf{\eta}_i,\mathbf{\eta}_j)&(\mathbf{\eta}_i,\mathbf{\eta}_k)\\
(\mathbf{\eta}_j,\mathbf{\eta}_i)&(\mathbf{\eta}_j,\mathbf{\eta}_j)&(\mathbf{\eta}_j,\mathbf{\eta}_k)\\
(\mathbf{\eta}_k,\mathbf{\eta}_i)&(\mathbf{\eta}_k,\mathbf{\eta}_j)&(\mathbf{\eta}_k,\mathbf{\eta}_k)\\
\end{array}
\right].
\]

As shown in Fig. \ref{fig:hessian}, $d_i \mathbf{n}_i + d_j
\mathbf{n}_j + d_k \mathbf{n}_k = 0$,
\[
\mathbf{\eta}_i = \frac{\mathbf{n}_i}{rd_i}, \mathbf{\eta}_j =
\frac{\mathbf{n}_j}{rd_j}, \mathbf{\eta}_k =
\frac{\mathbf{n}_k}{rd_k},
\]
where $r$ is the radius of the incircle of the triangle. Suppose
$(x_i,x_j,x_k)\in\mathbb{R}^3$ is a vector in $\mathbb{R}^3$, then
\[
[x_i,x_j,x_k]  \left[
\begin{array}{ccc}
(\mathbf{\eta}_i,\mathbf{\eta}_i)&(\mathbf{\eta}_i,\mathbf{\eta}_j)&(\mathbf{\eta}_i,\mathbf{\eta}_k)\\
(\mathbf{\eta}_j,\mathbf{\eta}_i)&(\mathbf{\eta}_j,\mathbf{\eta}_j)&(\mathbf{\eta}_j,\mathbf{\eta}_k)\\
(\mathbf{\eta}_k,\mathbf{\eta}_i)&(\mathbf{\eta}_k,\mathbf{\eta}_j)&(\mathbf{\eta}_k,\mathbf{\eta}_k)\\
\end{array}
\right] \left[
\begin{array}{c}
x_i\\
x_j\\
x_k
\end{array}
\right] = \|x_i\mathbf{\eta}_i + x_j\mathbf{\eta}_j +
x_k\mathbf{\eta}_k\|^2 \ge 0.
\]
If the result is zero, then $(x_i,x_j,x_k) =
\lambda(u_i,u_j,u_k),\lambda \in \mathbb{R}$. That is the null space
of the Hessian matrix. In the admissible metric space $\Omega_u$,
$u_i+u_j+u_k=C (C=3)$, then $du_i+du_j+du_k=0$. If $(du_i,du_j,du_k)$ belongs to the null space, then $(du_i,du_j,du_k)=\lambda(u_i,u_j,u_k)$, therefore,
$\lambda(u_i + u_j + u_k)=0$. Because $u_i,u_j,u_k$ are positive,
$\lambda=0$. In summary, the energy on $\Omega_u$ is convex.
\end{proof}

\smallskip
\begin{theorem}
The mapping $\nabla E: \Omega_u \to \Omega_\theta, (u_i,u_j,u_k) \to
(\cot\theta_i,\cot\theta_j,\cot\theta_k)$ is a diffeomorphism.
\end{theorem}

\smallskip
\begin{proof}
The energy $E(u_i,u_j,u_k)$ is a convex function defined on the
convex domain $\Omega_u$, according to Lemma \ref{lem:embedding},
$\nabla E: (u_i,u_j,u_k) \to
(\cot\theta_i,\cot\theta_j,\cot\theta_k)$ is a diffeomorphism.
\end{proof}

\section{Euclidean Polyhedral Surface}
\label{sec:polyhedral}

\vspace{2mm}
In this section, we consider the whole polyhedral surface.

\vspace{-2mm}
\subsection{Closed Surfaces}
\label{sec:closed}

Given a polyhedral surface $(S,T,{\mathbf{d}})$, the admissible metric space
 and the edge weight have been defined in Section \ref{sec:discrete} respectively.

\smallskip
\begin{lemma}The admissible metric space $\Omega_u$ is convex.
\label{lem:convexity_mesh_metric_space}
\end{lemma}

\smallskip
\begin{proof}
For a triangle $\{i,j,k\}\in F$, define
\[
 \Omega_u^{ijk} :=
\{(u_i,u_j,u_k)|(\sqrt{u_i},\sqrt{u_j},\sqrt{u_k})\in E_{\mathbf{d}}(2)\}.
\]
Similar to the proof of Lemma \ref{lem:convexity_metric_space},
$\Omega_u^{ijk}$ is convex. The admissible metric space for the mesh
is
\[
    \Omega_u = \bigcap_{\{i,j,k\}\in F} \Omega_u^{ijk}\bigcap
    \{(u_1,u_2,\cdots,u_m)|\sum_{k=1}^m
    u_k = m\},
\]
the intersection $\Omega_u$ is still convex.
\end{proof}

\smallskip
\begin{definition} [{Differential Form}]
The differential form $\omega$ defined on $\Omega_u$ is the
summation of the differential form on each face,
\[
    \omega = \sum_{\{i,j,k\}\in F} \omega_{ijk} = \sum_{i=1}^m 2w_i du_i,
\]
where $\omega_{ijk}$ is given in Eqn. (\ref{eqn:1_form}) in
Corollary \ref{cor:closed_1_form}. $w_i$ is the edge weight on
$e_i$.
\end{definition}

\smallskip
\begin{lemma}The differential form $\omega$ is a closed 1-form.
\end{lemma}

\smallskip
\begin{proof}
According to Corollary \ref{cor:closed_1_form},
\[
    d\omega = \sum_{\{i,j,k\}\in F} d\omega_{ijk} = 0.
\]
\end{proof}

\smallskip
\begin{lemma}
The energy function
\[
E(u_1,u_2,\cdots, u_n) = \sum_{\{i,j,k\}\in F}
E_{ijk}(u_1,u_2,\cdots, u_n)= \int^{(u_1,u_2,\cdots,
u_n)}_{(1,1,\cdots,1)} \sum_{i=1}^n w_i du_i
\]
is well defined and convex on $\Omega_u$, where $E_{ijk}$ is the
energy on the face, defined in Eqn. (\ref{eqn:energy}).
\label{lem:convexity_mesh_energy}
\end{lemma}

\smallskip
\begin{proof}
For each face $\{i,j,k\}\in F$,  the Hessian matrices of $E_{ijk}$
is semi-positive definite, therefore, the Hessian matrix of the total
energy $E$ is semi-positive definite.

Similar to the proof of Lemma \ref{lem:convexity_energy}, the null
space of the Hessian matrix $H$ is
\[
    ker H = \{\lambda(d_1,d_2,\cdots, d_n),\lambda \in \mathbb{R}\}.
\]
The tangent space of $\Omega_u$ at $u=(u_1,u_2,\cdots, u_n)$ is denoted by
$T\Omega_u(u)$. Assume $(du_1,du_2,\cdots, du_n)\in T\Omega_u(u)$,
then  from $\sum_{i=1}^m u_i = m$, we get $\sum_{i=1}^m du_m = 0$.
Therefore,
\[
    T\Omega_u( u ) \cap Ker H = \{0\},
\]
hence $H$ is positive definite restricted on $T\Omega_u(u)$. So the total energy $E$ is convex on $\Omega_u$.
\end{proof}

\smallskip
\begin{theorem}
\label{thm:closed_surface} The mapping on a closed Euclidean
polyhedral surface $\nabla E: \Omega_u \to \mathbb{R}^m,
(u_1,u_2,\cdots, u_n) \to (w_1,w_2,\cdots, w_n)$ is a smooth
embedding.
\end{theorem}
\begin{proof}
The admissible metric space $\Omega_u$ is convex as shown in Lemma
\ref{lem:convexity_mesh_metric_space}, the total energy is convex as shown in Lemma \ref{lem:convexity_mesh_energy}. According to Lemma
\ref{lem:embedding}, $\nabla E$ is a smooth embedding.
\end{proof}

\subsection{Open Surfaces}
\label{sec:open}
By the double covering technique \cite{Gu03SGP}, we can convert a polyhedral
surface with boundaries to a closed surface. First, let
$(\bar{S},\bar{T})$ be a copy of $(S,T)$, then we reverse the
orientation of each face in $\bar{M}$, and glue two surfaces $S$ and
$\bar{S}$ along their corresponding boundary edges, the resulting
triangulated surface is a closed one. We get the following corollary

\smallskip
\begin{corollary}
\label{cor:open_surface} The mapping on an Euclidean polyhedral
surface with boundaries $\nabla E: \Omega_u \to$ $\mathbb{R}^m,
(u_1,u_2$,\\$\cdots, u_n)$$ \to $$(w_1,w_2,\cdots, w_n)$ is a smooth
embedding.
\end{corollary}

\smallskip
Surely, the cotangent edge weights can be uniquely obtained from the
discrete heat kernel. By combining Theorem \ref{thm:closed_surface}
and Corollary \ref{cor:open_surface}, we obtain the major Theorem \ref{thm:main}, \emph{Global Rigidity Theorem}, of this work.

\section{Numerical Experiments}
\label{sec:experiment}
From above theoretic deduction, we can design the algorithm to
compute discrete metric with user prescribed edge weights.

\smallskip
\begin{problem}
Let $(S,T)$ be a triangulated surface,
$\mathbf{\bar{w}}(\bar{w}_1,\bar{w}_2,\cdots, \bar{w}_n)$ are the
user prescribed edge weights. The problem is to find a discrete
metric $\mathbf{u}=(u_1,u_2,\cdots, u_n)$, such that this metric
$\mathbf{\bar{u}}$ induces the desired edge weight $\mathbf{w}$.
\end{problem}

The algorithm is based on the following theorem.

\smallskip
\begin{theorem} Suppose $(S,T)$ is a triangulated surface. If there
exists an $\mathbf{\bar{u}}\in \Omega_u$, which induces
$\mathbf{\bar{w}}$, then $\mathbf{u}$ is the unique global minimum
of the energy
\begin{equation}
E(\mathbf{u}) = \int_{(1,1,\cdots,1)}^{(u_1,u_2,\cdots,u_n)}
\sum_{i=1}^n (\bar{w}_i - w_i) d\mu_i. \label{eqn:algorithm_energy}
\end{equation}
\end{theorem}

\smallskip
\begin{proof}
The gradient of the energy $\nabla E(\mathbf{u}) = \bar{\mathbf{w}}
- \mathbf{w}$, and since $\nabla E(\mathbf{\bar{u}})=0$, therefore
$\mathbf{\bar{u}}$ is a critical point. The Hessian matrix of
$E(\mathbf{u})$ is positive definite, the domain $\Omega_u$ is
convex, therefore $\mathbf{\bar{u}}$ is the unique global minimum of
the energy.
\end{proof}

\begin{figure*}
\centering
\begin{tabular}{ccc}
\includegraphics[height=1.6in]{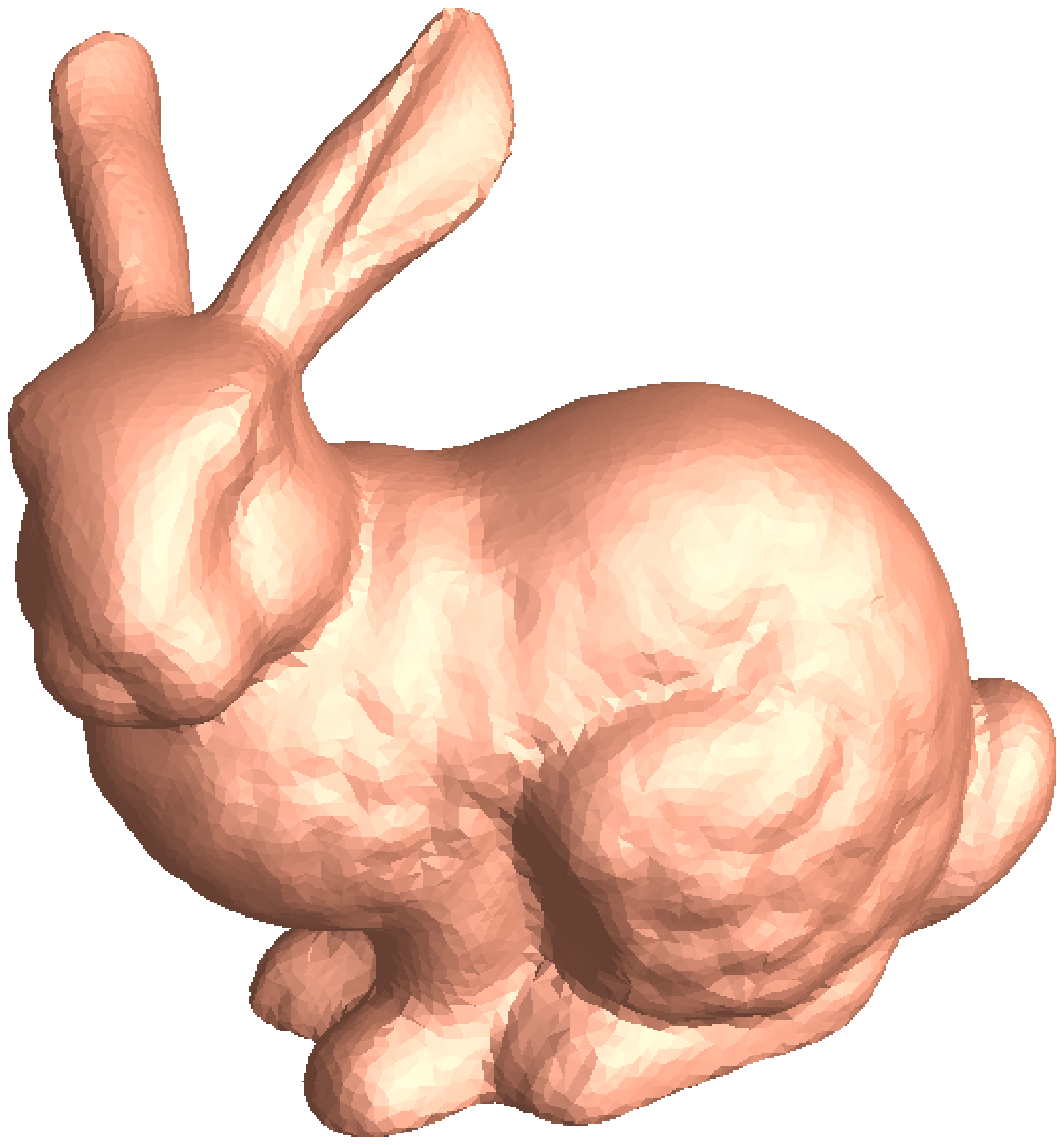} &\includegraphics[height=1.6in]{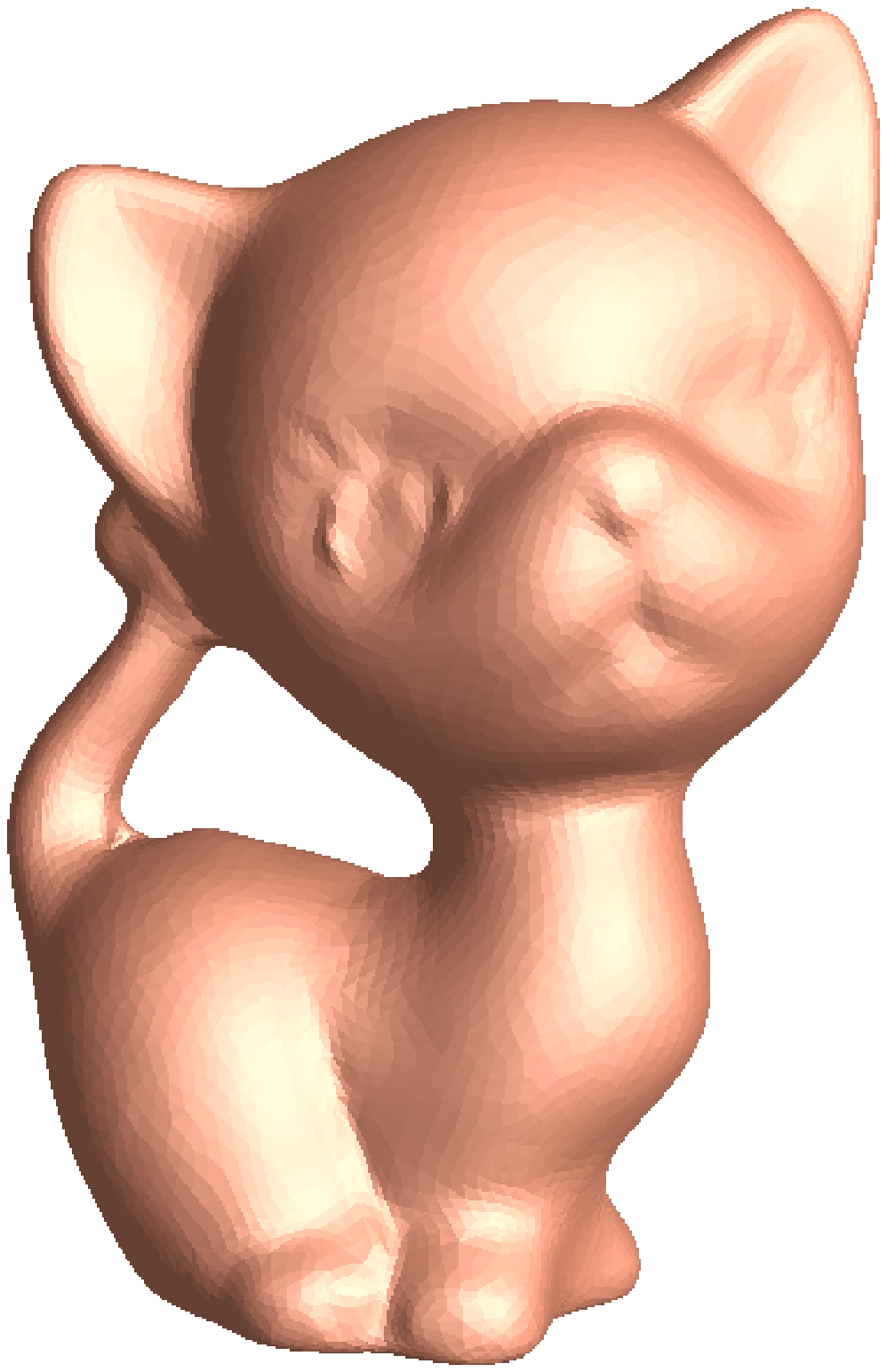} &
\includegraphics[height=1.6in]{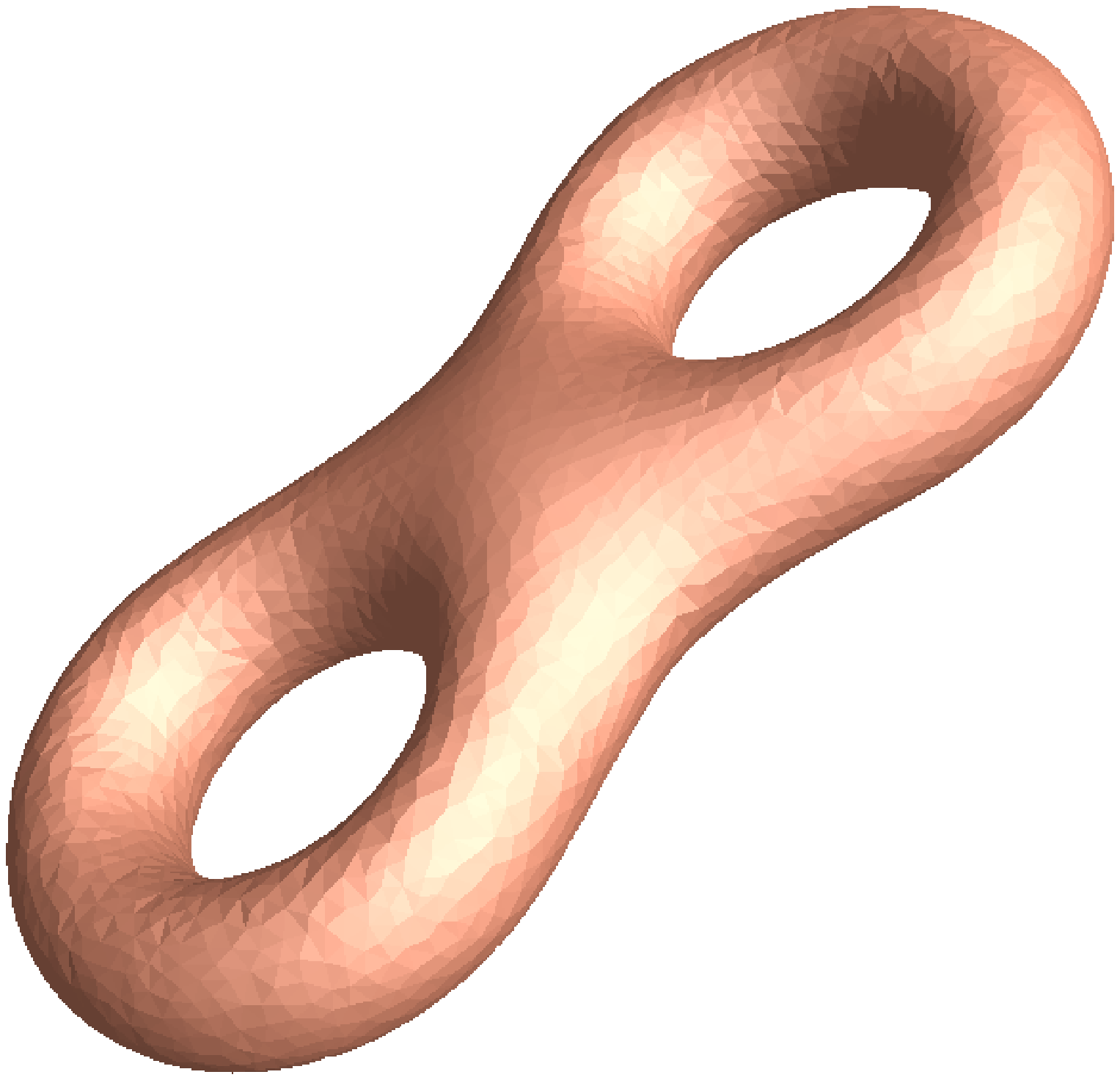}\\
\includegraphics[height=1.6in]{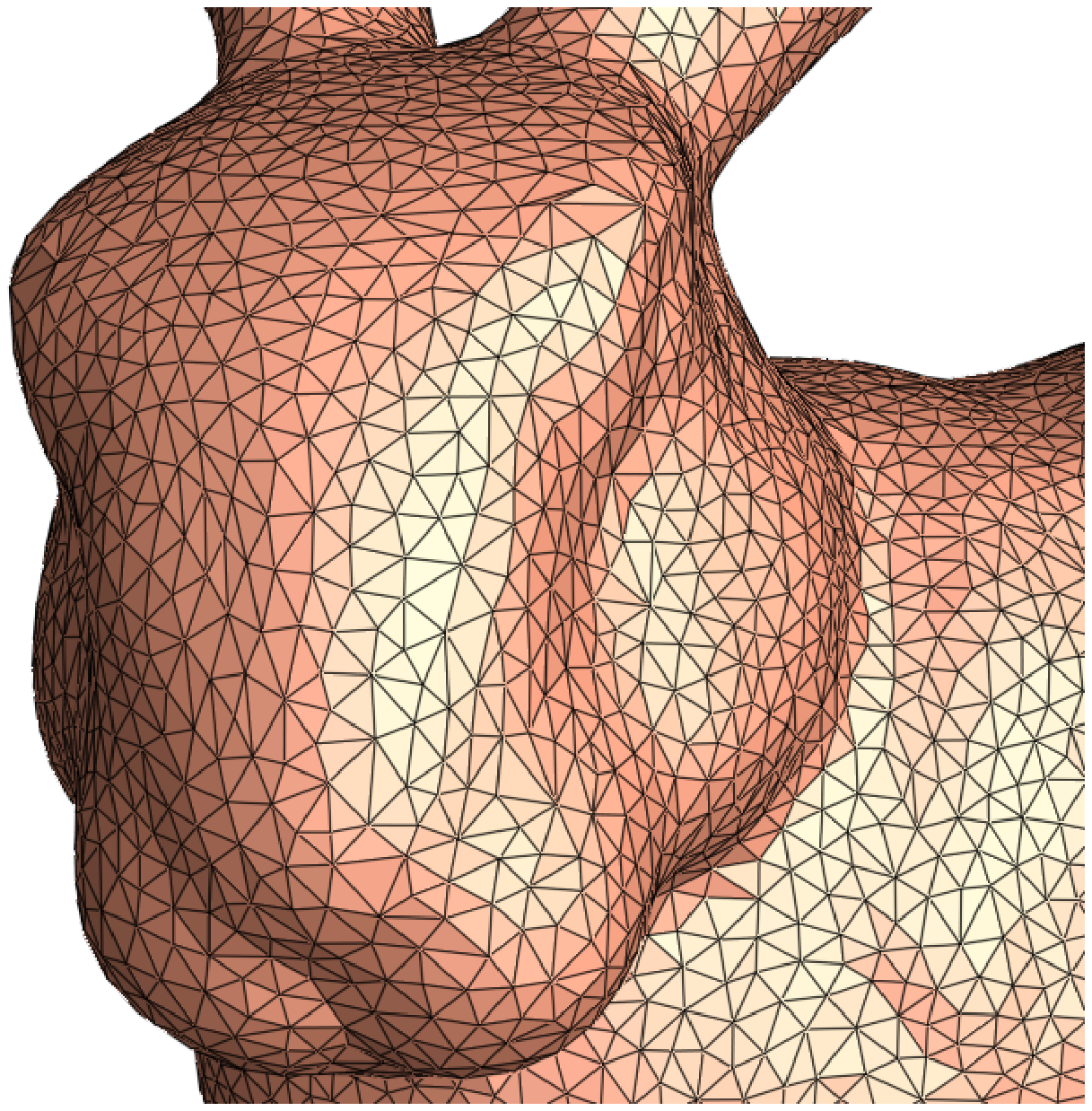} &
\includegraphics[height=1.6in]{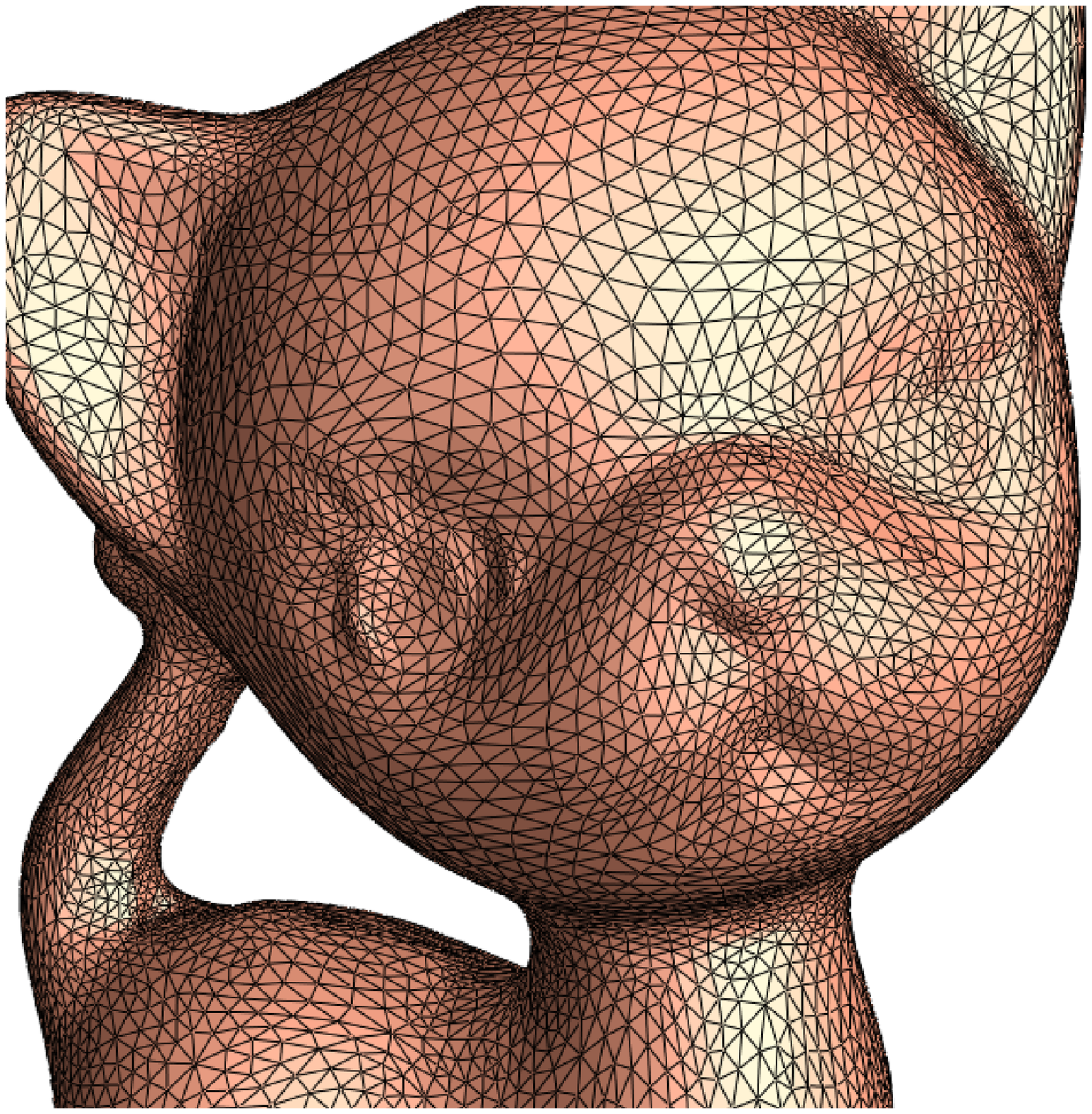}&
\includegraphics[height=1.6in]{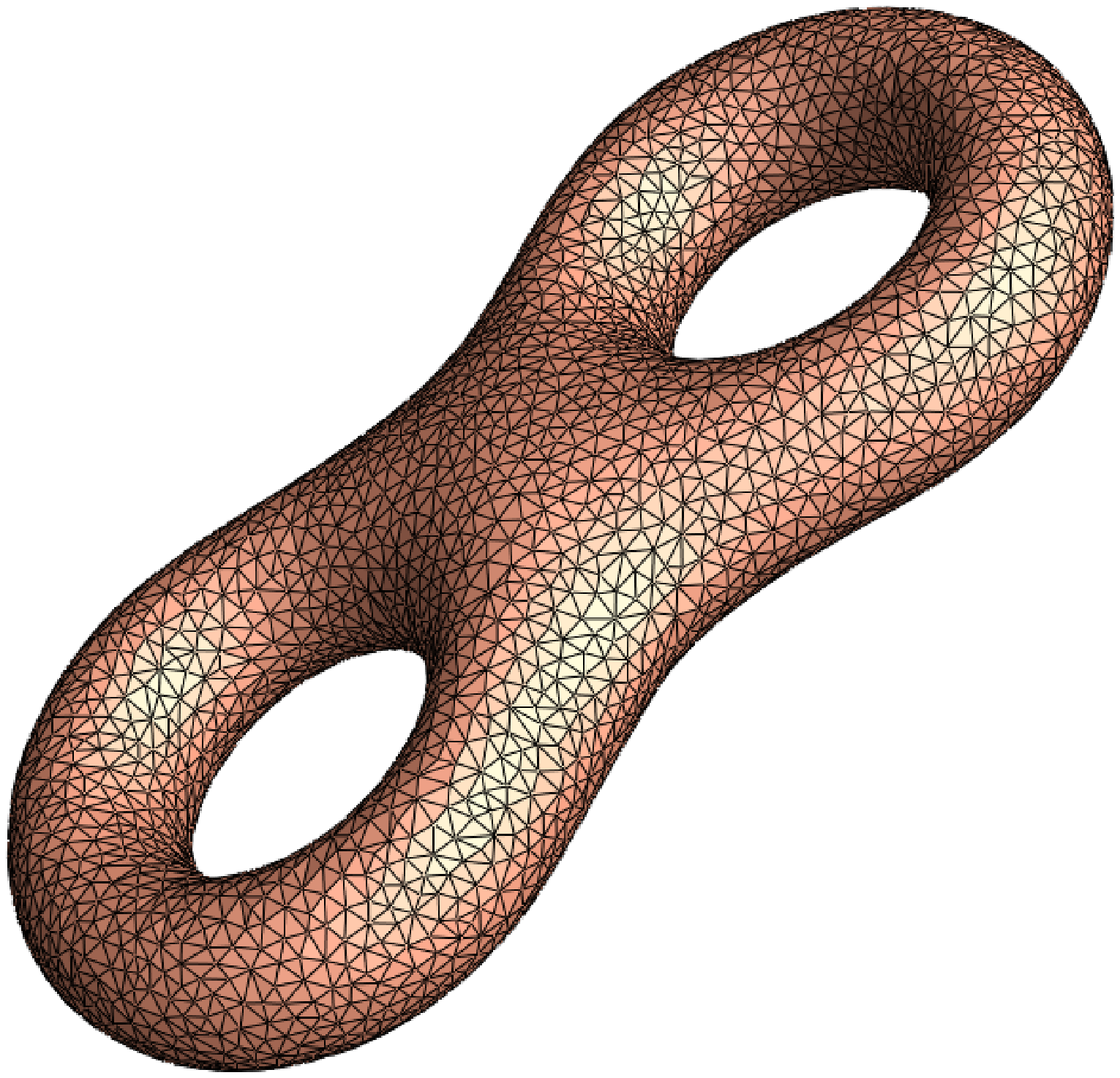}\\
Genus $0$ &Genus $1$&Genus $2$\\
\end{tabular}
\caption{Euclidean polyhedral surfaces used in the experiments.}
\label{fig:meshes}
\end{figure*}

In our numerical experiments, as shown in Fig. \ref{fig:meshes}, we tested surfaces with different
topologies, with different genus, with or without boundaries. All
discrete polyhedral surfaces are triangle meshes scanned from real
objects. Because the meshes are embedded in $\mathbb{R}^3$, they
have induced Euclidean metric, which are used as the desired metric
$\mathbf{\bar{u}}$. From the induced Euclidean metric, the desired
edge weight $\mathbf{\bar{w}}$ can be directly computed. Then we set
the initial discrete metric to be the constant metric
$(1,1,\cdots,1)$. By optimizing the energy in Eqn. (\ref{eqn:algorithm_energy}), we can reach the global minimum, and
recovered the desired metric, which differs from the induced
Euclidean metric by a scaling.

\section{Future Work}
\label{sec:future}
 We conjecture that the Main Theorem
\ref{thm:main} holds for arbitrary dimensional Euclidean polyhedral
manifolds, that means discrete Laplace-Beltrami operator (or
equivalently the discrete heat kernel) and the discrete metric
for any dimensional Euclidean polyhedral manifold are mutually
determined by each other. On the other hand, we will explore the
possibility to establish the same theorem for different types of
discrete Laplace-Beltrami operators.

\bibliographystyle{abbrv}

\end{document}